# DAMPING RATE LIMITATIONS FOR TRANSVERSE DAMPERS IN LARGE HADRON COLLIDERS *

V. A. Lebedev †, Fermilab, Batavia, USA

*Abstract*

The paper focuses on two issues important for design and operation of bunch-by-bunch transverse damper in a very large hadron collider, where fast damping is required to suppress beam instabilities and noise induced emittance growth. The first issue is associated with kick variation along a bunch which affects the damping of head-tail modes. The second issue is associated with effect of damper noise on the instability threshold.

## INTRODUCTION

An achievement of maximum luminosity in a collider requires large beam current and small emittance. In hadron colliders of very large energy the collider size becomes so large that the frequency of lowest betatron sideband approaches kHz range where spectral density of acoustic and magnetic field noise is unacceptably large. This noise drives the emittance growth resulting in fast luminosity decay. Effective suppression of this emittance growth may be achieved by fast transverse damping [1,2]. Fast emittance growth and its suppression by the damper was demonstrated at the LHC commissioning [3,4]. The required damper gain grows with the size of the collider and approaches few turns for a collider which will follow the LHC (like FCC). The instability suppression is typically less demanding to damping rate but still it is another important reason for fast damping.

There are many phenomena which limit the maximum damper gain [5]. Here we discuss two of them in details.
(1) A damper gain increase results in a better suppression of zero head-tail mode. However, such increase may excite higher head-tail modes, and thus make the bunch unstable. This effect is exacerbated by presence of non-zero chromaticity and wake-fields which destroy symmetry of head-tail motion. As will be seen below an introduction of kick non-uniformity along the bunch may allow significant reduction of excitation of head-tail modes and, consequently, increases the beam stability margin.
(2) Any practical damper has internal noise. Depending on damper design it is related to the thermal noise of its preamps and/or noise of digitization. This noise drives small amplitude beam motion which due to betatron frequency spread results in an emittance growth. The betatron motion non-linearity introduced for suppression of head-tail modes makes this noise-induced diffusion depending on a particle betatron amplitude. With time that changes the particle transverse distribution and, consequently, may result in a loss of Landau damping. This phenomenon was observed in the LHC where the beam could lose transverse stability minutes after bringing the beams to collisions without any visible changes in the machine. The effect was pronounced stronger in the case of external excitation of transverse motion [6,7]. The beam stability study based on the multiparticle tracking is reported in Ref. [7]. It showed that the latency of stability loss is related to changes in the distribution function induced by the damper noise. In this paper we consider a semi-analytical theory which attempts to show details of the process in a one-dimensional model.

Below we assume that the damper is bunch-by-bunch type so that each bunch is damped separately.

## DAMPING OF INTRABUNCH MOTION

For analysis of intrabunch motion we use the air-bag square-well (ABS) model initially suggested in Ref. [8] and actively used by A. Burov for analysis of bunch damping (see for example [9]).

In this model the bunch is presented by two fluxes moving in opposite directions with particle reflection at the bucket boundaries. In difference to the linear longitudinal motion in the air-bag model [10] where the bunch density is picked at the bunch ends this model has a uniform density distribution along bunch. Therefore, ABS model better suits for description of damper effect on damping of head-tail modes.

In dimensionless variables the equations of motion for two fluxes are:

$$\frac{\partial x_1}{\partial \tau} + \frac{\partial x_1}{\partial s} + \frac{\chi}{i} x_1 = \frac{f}{2i} + \frac{q}{2i}(x_1 - x_2), \quad (1)$$
$$\frac{\partial x_2}{\partial \tau} - \frac{\partial x_2}{\partial s} - \frac{\chi}{i} x_2 = \frac{f}{2i} + \frac{q}{2i}(x_2 - x_1).$$

where $x_1$ and $x_2$ are the transverse coordinates for the respective fluxes, $\chi = (\xi/\nu_s)(\Delta p/p)$ is the head-tail phase, $\xi$ is the tune chromaticity, $\nu_s$ is the synchrotron tune, $\pm\Delta p/p$ represent the momentum deviations for particles in the positive and negative fluxes, $\tau = \omega_s t$ is the dimensionless time, $s \in [0, \pi]$ is the dimensionless longitudinal particle coordinate, $q = \Delta\nu_{sc}/\nu_s$ is the space charge parameter, $\Delta\nu_{sc}$ is the space charge tune shift, and $f$ characterizes the forces coming from the damper and wake-fields. Following Ref. [9] we introduce the new transverse coordinate:

$$x = \begin{cases} e^{i\chi s} x_1, & s = \psi, \quad 0 \le \psi \le \pi, \\ e^{-i\chi s} x_2, & s = -\psi, \quad -\pi \le \psi \le 0. \end{cases} \quad (2)$$

Here we also introduced the phase $\psi$ describing the synchrotron motion so that $s = |\psi|, \psi \in [-\pi, \pi]$. Performing substitutions we can reduce two equations in Eq. (1) to

___________________________________________
* Work supported by Fermi Research Alliance, LLC under Contract No. De-AC02-07CH11359 with the United States Department of Energy
† val@fnal.gov

one:
$$\frac{\partial x}{\partial \tau} + \frac{\partial x}{\partial \psi} = \frac{1}{2i}\left(fe^{-i\chi s} + q(x(\psi) - x(-\psi))\right). \quad (3)$$

The force coming from the wake is determined by the following equation:
$$f(s) = \int_0^\pi W(s'-s)(x_1(s') + x_2(s'))ds'. \quad (4)$$

In this paper we consider two wake-functions: the constant wake –
$$W(s) = W_0 \theta(s), \quad (5)$$
and the resistive wall wake –
$$W(s) = \sqrt{\pi/4}\, W_0\, \theta(s)/\sqrt{s}. \quad (6)$$

The coefficient in the resistive wake definition was chosen so that for the uniform bunch displacement the force at the bunch tail would be equal for both wakes.

We assume that the force coming from the damper is determined by the following equation:
$$f(s) = -i\frac{G}{2\pi}\cos(k_k(s - \pi/2 + \phi_k)) \times$$
$$\int_0^\pi (x_1(s') + x_2(s'))\cos(k_p(s' - \pi/2 + \phi_p))ds'. \quad (7)$$

Here $k_p$ and $\phi_p$ determine the sensitivity of damper pickup to a particle position along the bunch, and $k_k$ and $\phi_k$ determine dependence of the kick on the longitudinal coordinate along the bunch.

In the absence of space charge, damping and wakes the solutions of Eq. (3) are:
$$x_n \equiv x_n(\tau,\psi) = a_n e^{in(\tau-\psi)}. \quad (8)$$

In the first order of perturbation theory when only a damper is present (no wakes and space charge) we obtain the growth rate:
$$\lambda_n \equiv \frac{1}{2|a_n|^2}\frac{d}{d\tau}|a_n|^2 = -\frac{G}{2}\text{Re}\left(R_n(k_p,\phi_p)\overline{R(k_k,\phi_k)}\right), \quad (9)$$
where
$$R_n(k_x,\phi_x) = \int_0^\pi e^{i\chi\psi}\cos\left(k_x\left(\psi - \frac{\pi}{2} + \phi_x\right)\right)\cos(n\psi)\frac{d\psi}{\pi}, \quad (10)$$
$$x = k, p.$$

As one can see from Eq. (9) all modes are damped (have negative growth rates) if $k_p = k_k$ and $\phi_p = \phi_k$.

In the general case we look for a solution in the form:
$$x_n = e^{\lambda_n \tau}\sum_{m=-N_m}^{N_m} A_{nm} e^{im\psi}. \quad (11)$$

where $N_m$ determines how many harmonics approximate the exact solution. Substituting this equation into Eq. (3), using definitions of Eqs. (4) and (7), multiplying obtained equation by $e^{-in\psi}$ and integrating we obtain a system of $2N_m+1$ linear equations. The eigen-values and eigen-vectors of this matrix equation yield complex frequencies for each mode and its structure ($x_n(\psi)$). To warrant a solution accuracy, the 161 modes (±80) were used. After finding the eigen-vectors the modes were ordered in ascending order of imaginary part of $\lambda_n$ (tune shift).

First, we consider the instability in the absence of damper and the space charge. Calculations show that for $\chi$ = 0 the transverse mode coupling instability threshold is: $W_0 = W_{th} \approx 0.363$ for the step-like wake and for $W_0 = W_{thr} \approx 0.383$ for the resistive wall wake. In further discussion we will characterize the wake strength relative to these thresholds.

Figure 1 shows dependencies of growth rates on mode frequencies for few lowest modes for the wake strengths twice above threshold, and for $\chi$ = 0 and $\chi$ = -2. For $\chi$ = 0 (strong head-tail case) and the wake twice above threshold only 0-th and 1-st modes are coupled making only one mode unstable. As one can see from the bottom plot many modes became unstable for $\chi$ = -2. Although growth rates for both wakes (step-like and resistive wall) are close the tune shifts of the modes are significantly larger for the resistive wall wake.

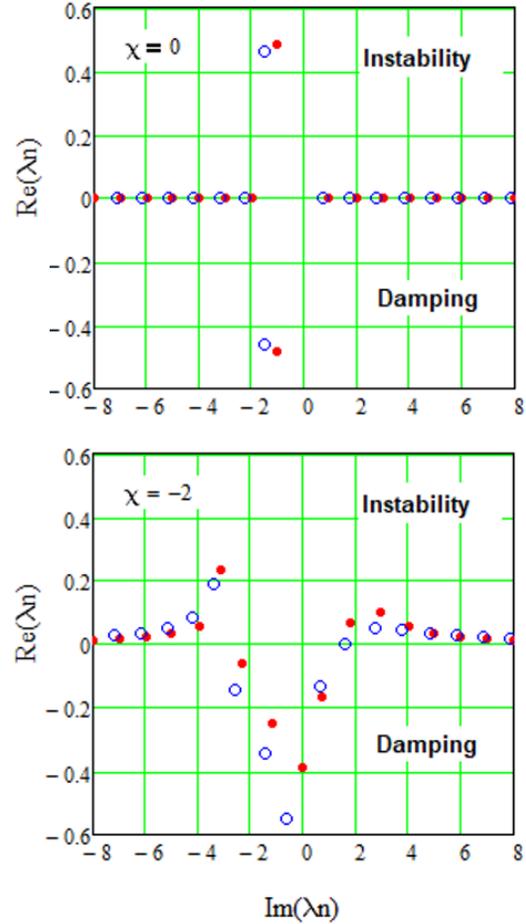

Figure 1: Dependence of growth rate, Re($\lambda_n$), on the mode coherent frequency, Im($\lambda_n$), for different modes and the wake amplitude twice above threshold; top – $\chi$ = 0, bottom – $\chi$ = -2; red dots – step-like wake, blue circles – resistive wall wake.

Further we characterize damping by the growth rate of the most unstable mode. Typically, it is the mode for which $n \approx \chi$. Figure 2 shows the growth rate of the most unstable mode on the head-tail phase, $\chi$, for different damper gains when both pickup and kicker have flat responses ($k_p = k_k = 0$). One can see in the top plot that there is no instability for $G = 0$ and $\chi = 0$ as should be expected below the instability threshold. However, for $G = 0$ the beam is unstable for any other (non-zero) head-tail phase. An increase of the damper gain reduces the growth rate for the most unstable mode everywhere except close vicinity of $\chi = 0$. Optimal damping is achieved at $G \approx 4$ where for the case twice below threshold the beam is stable for $\chi \in [0.5, 1.4]$ for both wakes. Further increase of the gain does not improve beam stability. For the wake twice above the threshold the beam is unstable for all $\chi$. Note that the considered model does not have Landau damping (discussed below) which stabilizes the beam if the growth rate is sufficiently small and these calculations do not show actual stability thresholds. Note also that the oscillations in the growth rates with $\chi$ are related to switching from one to another most unstable mode, so that one period represents the growth rate for one mode.

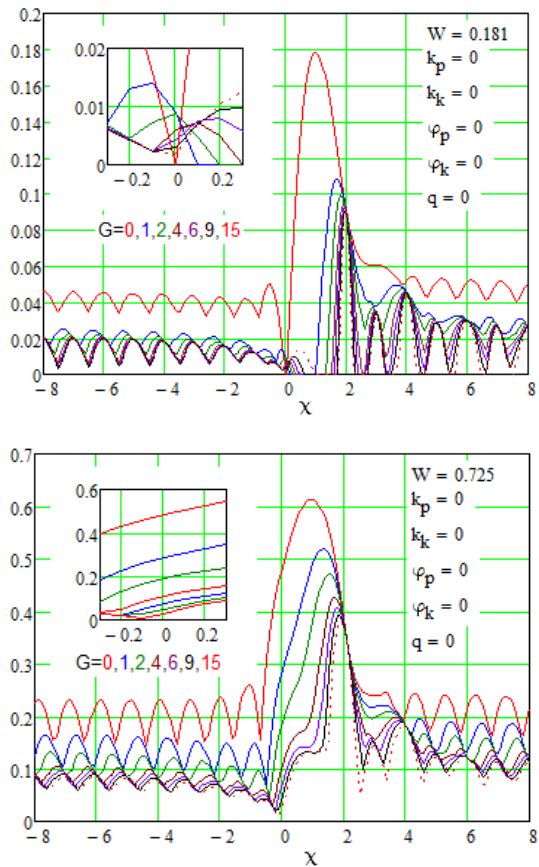

Figure 2: Dependence of the growth rate of the most unstable mode on $\chi$ for different damper gains ($G = 0, 1, 2, 4, 6, 9, 15$) for wake amplitudes twice below (top) and twice above the threshold; the step-like wake. Insets show dependences near $\chi = 0$.

Now we consider how changes in the response functions of pickup and kicker affect the beam stability. Figure 3 presents dependences of the growth rate of the most unstable mode on $\chi$ for different damper responses. As one can see for negative $\chi$ an increase of $k_p = k_k$ from 0 to 1 reduces the growth rate of most unstable mode by about 2 times. One can also see from the top plot that there is an area near $\chi = 0$ where all modes are stable. Variations of $\phi_p$ and $\phi_k$ and making $k_p$ and $k_k$ different did not exhibit stability improvement.

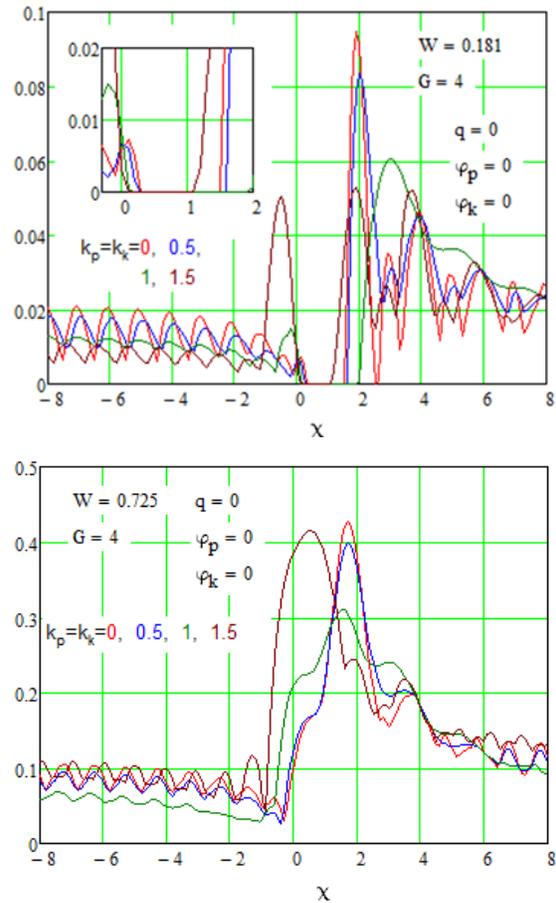

Figure 3: Dependence of the growth rate of the most unstable mode on $\chi$ for different damper responses for the cases of the beam intensity twice less (top) or twice more (bottom) than the strong head-tail threshold; the step-like wake.

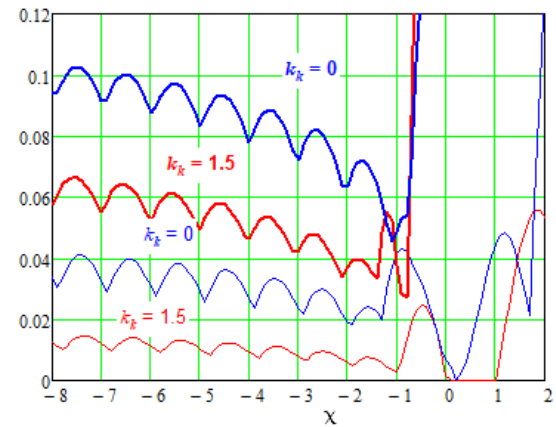

Figure 4: Dependence of the growth rate of the most unstable mode on $\chi$ for different kicker responses: red lines - $k_k = 1.5$, blue lines - $k_k = 0$; top two lines - $W$ is twice above threshold, bottom two lines - $W$ is twice below threshold; for all curves: $k_p = 1.5$, $\phi_p = \phi_k = q = 0$; the resistive wall wake.

All calculations were also repeated for the resistive wall wake and for different space charge parameter $q$. The results show that there is a reduction of the growth rate of most unstable mode by about two times for $k_p = k_k \approx 1$ in comparison with $k_p = k_k = 0$. Similar improvement happens in transition from $k_p \approx 1$, $k_k = 0$ to $k_p = k_k \approx 1$.

In the present LHC damper the pickup response to particle position is harmonic at 400 MHz frequency. The bunch length of 18 cm (~2σ) corresponds $k_p \approx 1.5$. That is already close to the optimum. However, the present kicker response is flat ($k_k = 0$) and as can be seen in Figure 4 that negatively affects the beam stability. Thus, making the kicker waveform as a few-periods 400 MHz sinusoid (short enough to avoid overlapping of signals of different bunches) would reduce the excitation of head-tail modes by factor of ~2.

## EFFECT OF DAMPER NOISE ON THE INSTABILITY THRESHOLD

For a continuous beam and the smooth lattice approximation the equation of a particle motion under external force $F(t) = F_\omega e^{-i\omega t}$ is:

$$-\omega^2 x_i + \omega_0^2 \left(Q_0 + \Delta Q_{lat_i}\right)^2 x_i = -2\omega_0^2 Q_0 \left(\Delta Q_c \bar{x} - F_\omega\right). \quad (12)$$

Here $i$ enumerates particles, $\omega_0$ is the circular frequency of particle revolution, $Q_0$ is the small amplitude betatron tune, $\Delta Q_{lat_i} \equiv \Delta Q_{lat}(J_{x_i}, J_{y_i})$ is the tune shift of particle betatron motion due to lattice non-linearity for a particle with betatron actions $J_{x_i}$ and $J_{y_i}$, $\Delta Q_c = \Delta Q_{cw} - ig/(4\pi)$ is the coherent tune shift which includes the tune shifts due to ring impedance, $\Delta Q_{cw}$, and due to transverse damper with damping rate per turn equal to $g/2$. Following the standard recipe [10, 11] we obtain the beam response to an external perturbation:

$$\frac{\bar{x}_\omega}{F_\omega} = \frac{R(\delta\omega)}{\varepsilon(\delta\omega)}. \quad (13)$$

Here $\bar{x}_\omega$ is the Fourier harmonic of beam centroid determined as $\bar{x}(t) = \sum_{i=1}^{N} x_i(t)/N$,

$$R(\delta\omega) = \int_0^\infty \frac{\partial f}{\partial J_x} \frac{J_x dJ_x dJ_y}{\left(\delta\omega/\omega_0 - \Delta Q_{lat}(J_x, J_y) + i0\right)} \quad (14)$$

is the response function in the absence of particle interaction, $f = f(J_x, J_y)$ is the particle distribution function normalized so that $\int f(J_x, J_y) dJ_x dJ_y = 1$,

$$\varepsilon(\delta\omega) = 1 + \Delta Q_c R(\delta\omega) \quad (15)$$

is the beam permeability, $\delta\omega = \omega - \omega_n$ is the frequency deviation from $n$-th betatron sideband, $\omega_n = (n - Q_0)\omega_0$, $i0$ determines the rule of pole traversing, and we assume that a frequency shift with particle momentum is much smaller than the shift due to betatron motion non-linearity. That allowed us to omit an integration over momentum distribution in Eq. (14).

With minor corrections these formulas are also justified for a bunched beam in the weak head-tail approximation [12]. First, in addition to the betatron sidebands we need to account the synchro-betatron sidebands. That yields the resonant frequencies to be $\omega_{nm} = (Q_0 + n + mQ_s)\omega_0$, where $Q_s$ is the synchrotron tune. Second, we need to account that a damper kick may excite multiple synchrotron-betatron modes. That is accounted by coefficients $w_m$. Consequently, Eq. (13) is modified to the following form:

$$\bar{x}_{\omega_{nm}} = w_m \frac{R(\delta\omega_{nm})}{\varepsilon_{nm}(\delta\omega_{nm})} F_\omega. \quad (16)$$

Here $\delta\omega_{nm} = \omega - \omega_{nm}$, and in Eq. (15) we need to account that the coherent tune shifts are different for each mode $\Delta Q_c \to \Delta Q_{c_{nm}}$ so that:

$$\varepsilon_{nm}(\delta\omega_{nm}) = 1 + \Delta Q_{c_{nm}} R(\delta\omega_{nm}), \quad (17)$$

where $R(\delta\omega)$ is still determined by Eq. (14).

Eq. (16) determines the amplitude of particle motion for a given synchro-betatron mode. For small amplitude excitation each synchro-betatron mode is excited independently and to obtain the total motion in the bunch one needs to sum motions of all modes.

The instability boundary (*i.e.* maximum coherent tune shift $\Delta Q_{c_{nm}}$ for a given mode is determined by the condition when with growth $\Delta Q_{c_{nm}}$ the beam permeability approaches zero the first time at any possible detuning. That corresponds to the solution of equation,

$$\varepsilon_{nm}(\delta\omega) = 0, \quad (18)$$

for real $\delta\omega$, which determines the stability boundary in the complex plane of $\Delta Q_c$. As follows from Eq. (13) the beam response of stable beam for a given mode is amplified by $1/|\varepsilon_{nm}(\delta\omega_{nm})|$ times.

Damper noise drives the transverse beam motion which due to spread in the betatron tunes results in an emittance growth. In the absence of particle interaction and active damping the emittance growth rate is [1]:

$$\left(\frac{d\varepsilon}{dt}\right)_0 = \frac{\omega_0^2 \beta_{kick}}{4\pi} \sum_{n=-\infty}^{\infty} P_\theta(\omega_n), \quad (19)$$

where $\beta_{kick}$ is the horizontal beta-function at the kicker location, and $P_\theta(\omega)$ is the spectral density of kicker angular noise normalized so that the rms value of the kicks is:

$$\overline{\theta^2} = \int_{-\infty}^{\infty} P_\theta(\omega) d\omega.$$

Taking Eq. (16) into account we can rewrite Eq. (19) in the following form:

$$\frac{d\varepsilon}{dt} = \int_0^\infty D(J_x, J_y) f(J_x, J_y) dJ_x dJ_y. \quad (20)$$

Here

$$D(J_x, J_y) = \frac{\omega_0^2 \beta_{kick}}{4\pi} \sum_{n,m=-\infty}^{\infty} \frac{w_m^2 P_\theta(\omega_n)}{\left|\varepsilon_{nm}(\omega_0 \Delta Q_{lat}(J_x, J_y))\right|^2}, \quad (21)$$

and we accounted that the spectral density of kicker noise does not change across one synchro-betatron sideband, noises at different frequencies do not correlate, and only resonant frequencies drive the emittance growth.

It is straightforward to find the emittance growth for the case of zero chromaticity, when only zero's synchro-betatron mode is excited. Assuming strong damping,

$g_n \gg 4\pi \max(|\Delta Q_{cw}|, \sqrt{\overline{\Delta v^2}})$, octupole non-linearity in the horizontal plane only, $\Delta Q_{lat}(J_x) = a_{xx}J_x$, and Gaussian distribution, $f(J_x) = e^{-J_x/J_a} / J_a$, we obtain:

$$D(J_x) = \frac{\omega_0^2 \beta_{kick}}{4\pi} \sum_{n=-\infty}^{\infty} \frac{16\pi^2 \overline{\Delta v^2} P_\theta(\omega_n)}{g_n^2 \left|\int_0^\infty \frac{x e^{-x} dx}{x-y-i0}\right|^2}, \quad y = \frac{J_x}{J_a}. \quad (22)$$

Here $\sqrt{\overline{\Delta v^2}} = a_{xx}J_a$ is the rms frequency tune spread, and $g_n$ is the damper gain at the $n$-th betatron sideband. Substituting diffusion of Eq. (22) into Eq. (20) and performing numerical integration one obtains a perfect coincidence with the result obtained in Ref. [1]:

$$\frac{d\varepsilon}{dt} = \frac{\omega_0^2 \beta_{kick}}{4\pi} \sum_{n=-\infty}^{\infty} \frac{16\pi^2 \overline{\Delta v^2}}{g_n^2} P_\theta(\omega_n), \quad g_n \gg 4\pi\sqrt{\overline{\Delta v^2}}. \quad (23)$$

Note that Eq. (21) is applicable in the general case while Eq. (23) in the case of zero chromaticity and far away from the instability threshold. Note also that the derivation of Eq. (23) in Ref. [1] does not actually determine the tune relative to which $\sqrt{\overline{\Delta v^2}}$ is computed. This question is addressed by Eq. (21).

To find a change in the instability threshold related to a change in the distribution we need to investigate the distribution function evolution. Considering that the kicks are small and uncorrelated; and, consequently, the process is very slow relative to the betatron motion the evolution can be described by the diffusion equation. In the general case of uncoupled betatron motion the diffusion in the 2D-space of actions is described by the following diffusion equation:

$$\frac{\partial f}{\partial t} = \frac{\partial}{\partial J_x}\left(J_x D_x(J_x, J_y)\frac{\partial f}{\partial J_x}\right) + \frac{\partial}{\partial J_y}\left(J_y D_y(J_x, J_y)\frac{\partial f}{\partial J_y}\right). \quad (24)$$

Here the diffusion in the horizontal plane is determined by Eq. (21). The vertical plane diffusion is obtained by changing corresponding indices.

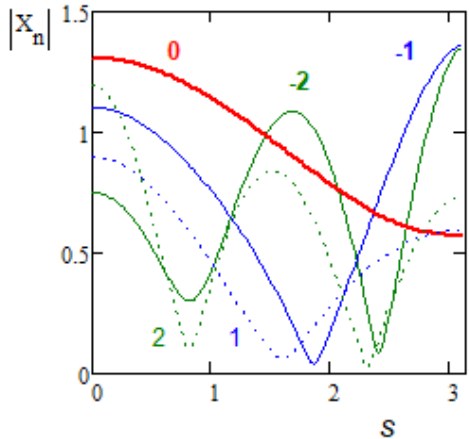

Figure 5: Dependencies of mode magnitudes, $|X_n| \equiv |x_{1n} + x_{2n}|$, along the bunch for the parameters of the LHC damper: $\chi = 1$, $k_p = 1.5$, $k_k = \phi_p = \phi_k = q = 0$, $W = 2W_{thr}$ for the resistive wall wake. Numbers show the mode numbers.

In the presence of impedance and chromaticity each kicker kick excites multiple head-tail modes. Only few of them are damped by the damper. Figure 5 shows shapes of few lowest head-tail modes for the damper model described in the previous section for the LHC parameters. As one can see all of them have significant variations along the bunch while the kicker kick is the same for all particles. Therefore, each kick in addition to the zero mode excites other modes. To find corresponding contributions we equalize the kick dependence along the bunch and weighted sum of the mode amplitudes:

$$\cos\left(k_k\left(|\psi| - \pi/2 + \phi_k\right)\right) = i\sum_m \hat{w}_m x_m(\psi). \quad (25)$$

where $x_m(\psi)$ is determined by Eq. (11) and is additionally normalized so that $x_m(\pi/2) = 1$. The solution of this equation yields coefficients $\hat{w}_m$. To obtain coefficients $w_m$ which determine relative excitation for different head-tail modes we additionally need to account how a given mode with amplitude $\hat{w}_m$ contributes to the emittance growth. That yields: $w_m^2 = \overline{|x_m(\psi)|^2} \hat{w}_m^2$. Figure 6 shows $\hat{w}_m$ for the modes presented in Figure 5. One can see that the mode zero has the largest contribution, $\hat{w}_0$, and the only one which has significant damping.

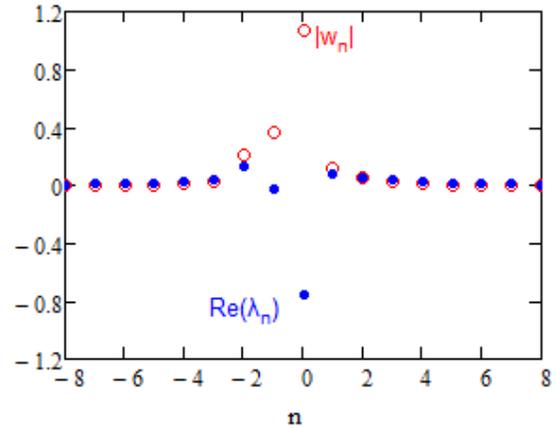

Figure 6: Dependences on the head-tail mode number for $|\hat{w}_n|$ (red circles) and the damping rate (blue dots).

To demonstrate an effect of damper noise on the beam stability boundary we initially assume that only one of the head-tail modes is near the threshold and it dominates the emittance growth. Applicability of this assumption we will discuss later. We also assume that the focusing non-linearity is in one plane only. That allows us to consider a one-dimensional problem. Then, from Eqs. (21) and (24) we obtain a simplified diffusion equation:

$$\frac{\partial f}{\partial \tau} = \frac{\partial}{\partial J_x}\left(J_x \hat{D}(J_x)\frac{\partial f}{\partial J_x}\right), \quad \hat{D}(J_x) = \frac{1}{\left|\varepsilon(\delta\omega(J_x))\right|^2}. \quad (26)$$

Here we transited to the dimensionless variables so that the action $J_x$ is measured in units of rms action $J_a$, and time $\tau$ is chosen to make the diffusion coefficient equal to the one in the absence of beam interaction. We also took into account that the diffusion is proportional $1/|\varepsilon|^2$ at the resonance frequency which is directly related to the action as $\Delta Q_{lat} = a_{xx} J_x$. That yields the univocal dependence of beam permeability on the action:

$$\varepsilon(\delta\omega(J_x)) = 1 + \frac{\Delta Q_c}{a_{xx}} \int_0^{J_{max}} \frac{\partial f}{\partial J'_x} \frac{J'_x dJ'_x}{J_x - J'_x + i0}. \quad (27)$$

where $J_{max}$ is determined by the ring acceptance.

The solution of Eq. (26) with beam permeability of Eq. (27) was carried out numerically. The action space was binned into boxes with boundaries at $J_n = n\Delta J$, $n \in [0, N_{max}]$, so that $f_n \Delta J$ is the probability to find a particle in $n$-th box and $f_n$ is the distribution function in the center of the box bounded by $J_n$ and $J_{n+1}$. An integration of Eq. (26) over $J$ through one box yields the particle flux through the boundary between boxes $n$ and $n+1$:

$$\Phi_{n+1} = J_{n+1} D(J_{n+1}) \frac{\partial f}{\partial J}\bigg|_{J=J_{n+1}} \rightarrow J_{n+1} D_{n+1} \frac{f_{n+1} - f_n}{\Delta J}. \quad (28)$$

Consequently, the change in the distribution is:

$$f_n(t_{k+1}) = f_n(t_k) + (\Phi_n(t_k) - \Phi_{n+1}(t_k))\Delta t, \quad \Delta t = t_{k+1} - t_k. \quad (29)$$

Time step $\Delta t$ was chosen so that to be well below the instability threshold of the difference scheme, which is determined by:

$$S \equiv \max_n (4 D_n J_n \Delta t / \Delta J^2) = 1.$$

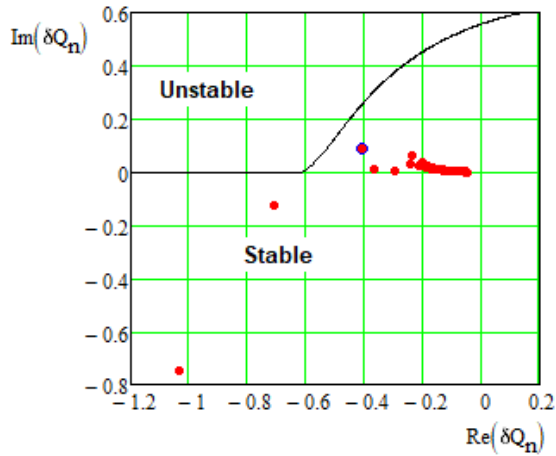

Figure 7: Ratios of coherent tune shits to the synchrotron tune for different modes for parameters of Figure 5. Black line presents the stability boundary for Gaussian beam with non-linearity parameter $a_{xx}$ chosen so that the most unstable mode (marked by blue circle) would be 25% below stability threshold.

For a harmonic perturbation $\delta f \cos(\kappa J)$ and $S \ll 1$ this difference scheme yields good approximation for small $\kappa$. However, it reduces damping at the highest frequency of $\kappa_{max} = \pi/(2\Delta J)$ by $(\pi/2)^2$ times. Note that a usage of implicit methods typically applied to the diffusion equation solving is limited by two circumstances. First, a computation of diffusion at any point in the action space uses the entire particle distribution and therefore computation of distribution at next point in time requires inversion of $N_{max} \times N_{max}$ matrix instead of three-diagonal matrix for the case of implicit scheme. Second, as will be shown below, the instability is developing at high frequencies. That requires small steps in time.

To accelerate computation of the integral in Eq. (27) it was reduced to a matrix multiplication so that the vector of beam permeability is equal to:

$$\boldsymbol{\varepsilon} = \mathbf{R}\mathbf{f}. \quad (30)$$

Here the vectors $\boldsymbol{\varepsilon} \equiv \varepsilon_n$ and $\mathbf{f} \equiv f_n$ determine the beam permeability and the distribution function. The elements of matrix $\mathbf{R}$ are determined by integration Eq. (27) between nearby actions $J_n$ using Tailor expansion of $f$. Numerical tests verified that Eq. (30) results in good approximation of integral (27) in the absence of discontinuities in the distribution.

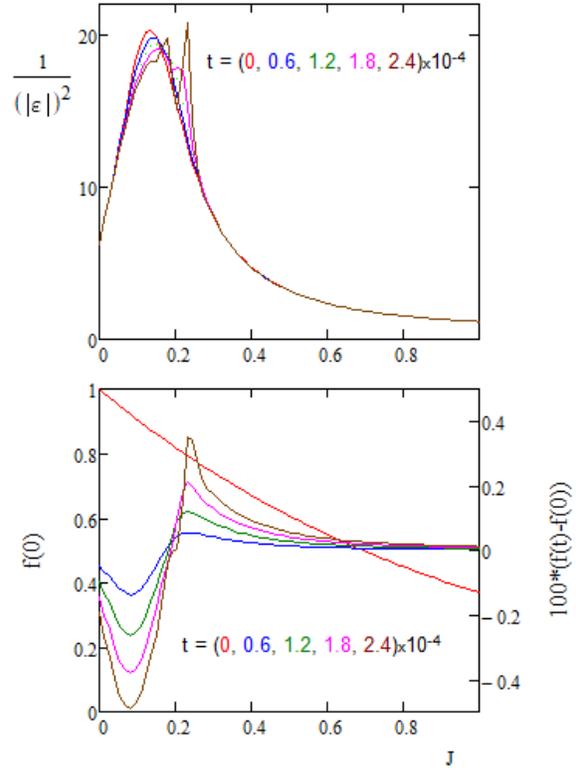

Figure 8: Dependence of dimensionless diffusion (top) and distribution function (bottom) on the action for different times, $t$; $a_{xx} = 0.02$, $\Delta Q_c = (-12.6+3.1i)10^{-3}$. Red curve in the bottom plot shows the initial distribution (left scale) and other curves changes of the distribution multiplied by 100 (right scale).

Simulations showed that loss of stability due to distribution evolution under kicker noise strongly depends on the phase of the coherent tune $r = \arg(\delta Q_n)$. Figure 7 presents the dimensionless coherent tune shifts (ratio of coherent tune shifts to the synchrotron tune) for different head tail modes for the parameters of Figure 5. The stability boundary was chosen to be 25% above most unstable mode for which $r = 168°$ (Re($\delta Q_n$)/ Im($\delta Q_n$) = -4.7). The distance from the stability boundary to the next mode closest to the boundary is about twice larger, and consequently its effect on the diffusion is 4 times smaller. Figure 8 shows a typical example of the evolution for initially Gaussian distribution. The figure also shows the corre-sponding diffusion. The value of $\Delta Q_c/a_{xx}$ was chosen so that the beam would be 25% below instability threshold (see Figure 7). In all simulations (as well as in Figure 8) it has been clearly seen that the instability, if happens, develops at the highest possible wave-number determined by $\Delta J$. An increase of $N_{max}$ decreases $\Delta J$ and the span in the distribution where the instability is initially developed. However, the location of the instability position in the action did not depend on $N_{max}$.

To explain the results of the simulations we consider the following model. We assume that the instability is developed at a small area near the action $J_r$. In this area we look for a solution in the following form:
$$f(J_x,t) = f_0(J_x) + \delta f(t)\cos(\kappa J + \psi), \quad (31)$$
where we assume the wave-number, $\kappa$, being very large, and the perturbation $\delta f(t) \equiv \delta f$ to be much smaller than the initial distribution $f_0(J_x)$. A perturbation in the distribution results in a perturbation in the response function. Substituting the perturbation of Eq. (31) into Eq. (14) we obtain a perturbation of response function:
$$\delta R \equiv \delta R(J_x) = -\frac{i\kappa}{a_{xx}} \delta f \int_0^\infty \frac{\sin(\kappa J + \psi)JdJ}{J_x - J + i0}, \quad (32)$$
where we accounted that the resonance frequency is $\delta\omega = a_{xx} J_x$. For large $\kappa$ the major contribution to the integral comes from the area near $J_x$. That allows us to extend the integration to $-\infty$. Then, the integration becomes straight forward. It results in:
$$\delta R = \frac{\pi\kappa J_x}{a_{xx}} e^{i(\kappa J_x + \psi)} \delta f(t). \quad (33)$$
Using Eqs. (15) and (26), we obtain the diffusion:
$$\hat{D} = \frac{1}{|1 + \Delta Q_c(R + \delta R)|^2} \approx D_r + \delta D,$$
$$\delta D = -D_r \operatorname{Re}\left(\frac{2\pi\kappa J_r \Delta Q_c}{\varepsilon_r} e^{i(\kappa J_x + \psi)}\right)\delta f, \quad (34)$$
where $\varepsilon_r = 1 + \Delta Q_c R_r$ is the beam permeability for unperturbed beam computed at the resonant tune $\delta\omega/\omega_0 = J_r a_{xx}$, $D_r = 1/|\varepsilon_r|^2$ is the corresponding diffusion, and in obtaining the second equality we used the Tailor expansion and replaced $J_x$ by $J_r$ in the non-oscillating term. As one can see a harmonic perturbation of the distribution results in a harmonic perturbation of the diffusion.

Taking into account that we consider only small aria in the action space in vicinity of $J_r$ and very large wavenumber $\kappa$ (see the definition below) we can replace $J_x$ inside $\partial/\partial J_x$ in Eq. (26) by $J_r$. That yields:
$$\frac{\partial}{\partial\tau}(f_0 + \delta f) = J_r \frac{\partial}{\partial J_x}\left((D_r + \delta D)\frac{\partial}{\partial J_x}(f_0 + \delta f)\right). \quad (35)$$
Accounting that the unperturbed function satisfies the following equation:
$$\frac{\partial f_0}{\partial \tau} = J_r \frac{\partial}{\partial J_x}\left(D_r \frac{\partial f_0}{\partial J_x}\right) \quad (36)$$
and leaving only linear terms in Eq. (35) we obtain a linear differential equation for the perturbation

$$\frac{\partial \delta f}{\partial \tau} = J_r \frac{\partial}{\partial J_x}\left(D_r \frac{\partial \delta f}{\partial J_x} + \delta D \frac{\partial f_0}{\partial J_x}\right). \quad (37)$$
We look for a solution in the following form:
$$\delta f = \tilde{f} e^{-\lambda \tau} \cos(\kappa I - \nu\tau). \quad (38)$$
Substituting it into Eq. (37), assuming initial Gaussian distribution $f_0 = e^{-J_x}$, and using Eq. (34) we obtain the damping rate as a function of $J_r$:
$$\lambda = J_r D_0 \kappa^2 \left(1 + e^{-J_r}\left(\operatorname{Im}(A) - \operatorname{Re}(A)/\kappa\right)\right),$$
$$A = \frac{2\pi J_r \Delta Q_c}{\varepsilon_r a_{xx}}. \quad (39)$$
For large $\kappa$ the last term can be neglected. Thus, for the Gaussian distribution the stability area for given $\Delta Q_c$ is determined by following equation,
$$1 + \frac{2\pi J_r \Delta Q_c}{a_{xx}} e^{-J_r} \operatorname{Im}\left(\frac{\Delta Q_c}{\varepsilon_r}\right) \geq 0, \quad (40)$$

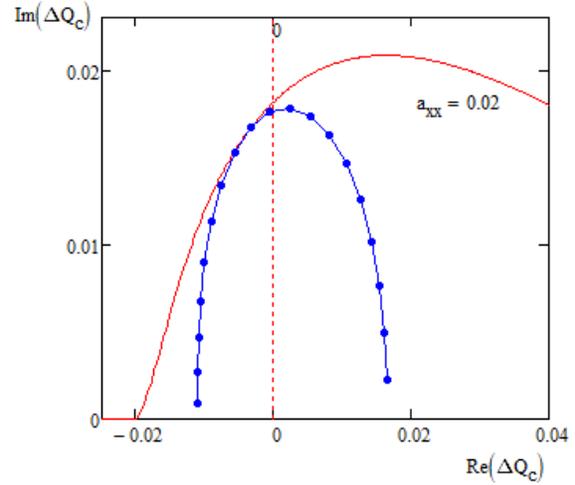

Figure 9: Stability diagram computed with accounting noise driven diffusion (blue curve) and without it (red curve.)

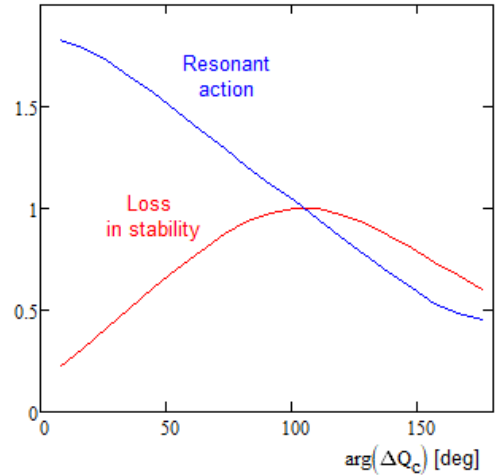

Figure 10: Dependence of the resonant action and the loss in stability on the angle of the coherent tune shift in the complex plane.

which must be satisfied for all $J_r$. Figure 9 presents the stability diagrams computed with the help of Eqs. (18) (red curve) and (40) (blue curve). One can see that the kicker noise results in significant reduction of the stability boundary. However, this reduction is negligible in vicinity of $\arg(\Delta Q_c) \approx 105°$. We will call the action $J_r$ at which the left-hand side in Eq. (40) approaches zero the first time the resonant action. It shows where instability develops when the beam is approaching to the instability boundary. Figure 10 shows how this resonant action depends on the angle of the coherent tune shift in the complex plane. The figure also shows the ratio of stability boundary sizes (ratio of $|\Delta Q_c|$ for given $r = \arg(|\Delta Q_c|)$ for curves presented in Figure 9). Numerical simulations verified the reduction of the stability boundary presented in Figure 9 and 10 and the location of the resonant action.

Taking into account that the considered above instability develops at high frequency and the resonant actions of different head-tail modes are different, we, in the first approximation, can neglect mutual interaction of different modes. That results in that the considered above model should be applicable to the situation when multiple modes are close to the instability boundary. If required it is straightforward to extend this model to multiple modes introducing summation of different modes in Eq. (34).

## CONCLUSIONS

An introduction of harmonic variation in the kicker waveform looks as a promising method for an increase of stability boundary for the LHC. Such a kicker does not work well for suppression of emittance growth due to injection errors. Therefore, the existing low frequency kicker should be retained and used for damping injection errors. A new kicker operating at 400 MHz base frequency could be used for the rest of the accelerating cycle and in the collisions. The power and space required for this new kicker are determined by the BPM noise and are well within the reach.

The considered above mechanism for reduction of the stability boundary points out underlying reasons behind the observations of transverse beam stability loss in the LHC. We need to note that in this model we neglected other diffusion mechanisms which affect the evolution of the distribution. In normal operating conditions the intra-beam scattering is the major diffusion mechanism. It counteracts the effects introduced by the damper noise and therefore a reduction of stability boundary due to kicker noise should be somewhat smaller. An additional noise used in the LHC experiments reduced relative effect of the IBS driven diffusion with subsequent reduction of the stability boundary observed in the experiments [6].

## ACKNOWLEDGMENTS

The author would like to thank A. Burov, E. Metral and X. Buffat for help in editing of this paper and many useful discussions on its subject.